\documentclass[showpacs,showkeys,amsmath,amssymb,aps,twocolumn,prl]{revtex4-1}
\usepackage{graphicx}
\usepackage{dcolumn}
\usepackage{bm}
\begin{document}

\preprint{}

\title{Fractional Brownian motors and Stochastic Resonance}

\author{Igor Goychuk}
 \email{igor.goychuk@physik.uni-augsburg.de}
\affiliation{Institute of Physics, University of Augsburg, Universit\"{a}tstr. 1, D-86135 Augsburg, Germany}

\author{Vasyl Kharchenko}
 \email{vasyl.kharchenko@physik.uni-augsburg.de}
\affiliation{Institute of Physics, University of Augsburg, Universit\"{a}tstr. 1, D-86135 Augsburg, Germany}
\affiliation{Institute of Applied Physics, 58 Petropavlovskaya str., 40030 Sumy, Ukraine}

\date{\today}

\begin{abstract}

We study fluctuating tilt Brownian ratchets based on fractional subdiffusion in  sticky viscoelastic media
characterized by a power law memory kernel. Unlike the normal diffusion case the rectification effect vanishes
in the adiabatically slow modulation limit and optimizes in a driving frequency range. It is shown also that
anomalous rectification effect is maximal  (stochastic resonance effect) at optimal temperature and can exhibit
a surprisingly good quality. Moreover,
subdiffusive current can flow in the counter-intuitive direction upon a change of temperature or driving
frequency. The dependence of anomalous transport on load exhibits a remarkably simple universality.    

\end{abstract}
\maketitle

\begin{figure}
 \includegraphics[width=0.7\columnwidth]{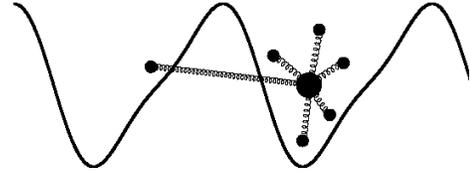}
 \caption{Ratchet potential and central Brownian particle coupled to auxiliary  Brownian particles 
 modeling viscoelastic environment.
 \label{Fig1}} 
\end{figure}

Such diverse research fields as anomalous diffusion and transport \cite{Hughes,Bouchaud,Metzler,WSBook}, 
Brownian ratchets \cite{Ajdari,Magnasco,Bartussek,ReimannRev,MarchesoniRev} and 
Stochastic Resonance (SR) \cite{SR} attracted much attention
over last two decades with a huge research literature produced and a number of insightful reviews written
which address both fundamental aspects of nonequilibrium
statistical physics and  various interdisciplinary 
applications in physics, chemistry, biology, and technology. The existing ratchet literature is restricted mostly to 
normal diffusion ratchets. Here, a rectification current can emerge for the particles diffusing in some periodic and
unbiased on average potential due to breaking the symmetry of thermal detailed balance by an external 
time-dependent driving. This in turn requires to break some spatiotemporal symmetry \cite{ReimannRev,MarchesoniRev}, for example, 
the spatial inversion symmetry as in Fig. \ref{Fig1} in the case of a fluctuating
tilt ratchet \cite{Magnasco,Bartussek} driven by harmonic force, which we consider in this work. 
The emergence of net directed motion in unbiased on average systems is
a strongly nonequilibrium and nonlinear effect, absent e.g. within the linear
response approximation, or linear Onsager regime of nonequilibrium thermodynamics.  
A characteristic feature of any
true ratchet is its ability to sustain a load, i.e. a force directed against  the transport direction. 
The presence of a non-zero stopping force allows to distinguish the genuine ratchets or Brownian motors 
capable to do useful work
from the futile ones or pseudo-ratchets \cite{ReimannRev,MarchesoniRev}. In any isothermal
Brownian motion which never ceases the dissipative loss of energy is compensated at thermal equilibrium by the 
energy (heat) gain due the thermal noise of 
environment so that on average the classical Brownian particle has kinetic energy $k_BT/2$ per degree of freedom. 
This ensures 
the absence of  a net directed motion and of the 
total heat exchange between the particle and its environment. The directed Brownian motion 
requires an external source of energy -- a part of it will be put into the directed motion and a part dissipated as an
\textit{excess} heat to the environment. 
The thermal noise plays in fact a constructive role here. First, as a sort of ``lubricant'' to smooth the 
friction and also providing
thermal energy fluctuations allowing to overcome potential barriers met on the particle's pathway. 
Restricting to the classical world, without noise the 
Brownian particle would remains localized in a potential well, starting there having
subthreshold energy and for a weak external driving.
Therefore, in such a setup 
one generally expects that the rectification current response to a subthreshold driving will increase 
with the noise intensity. However,
for a very strong noise the potential barriers cease to matter and one expects that rectification effect due to a
spatial asymmetry of potential 
will vanish. Therefore, there should exist an optimal thermal noise intensity, or temperature which typies Stochastic Resonance 
(SR) in broad sense. 
In the focus of this Letter are the ratchet and Stochastic Resonance effects in subdiffusive transport, i.e. mean displacement
grows sublinearly,
$\langle \delta x(t)\rangle \propto t^\alpha$, with $0<\alpha<1$, 
in a viscoelastic environment.

The very existence of subdiffusive ratchet transport is not obvious and can be questioned. 
For example, within a continuos time random walk (CTRW) mechanism of subdiffusion featured by divergent mean
residence times (MRTs) \cite{Hughes,Metzler} the current response to external periodic driving is asymptotically zero 
\cite{DOR}. 
This clearly prohibits any asymptotic rectification effect for such fluctuating tilt ratchets.  
 However, a subdiffusive rocked ratchet based on the fractional Brownian motion 
(fBm) \cite{Mandelbrot} does exist 
\cite{Goychuk10}, and we 
shall explain its unusual properties, showning in particular that the resonance nonadiabatic character of this anomalous 
ratchet effect is of the SR origin, and it is a genuine ratchet effect.

\textit{Theory.} The Fractional Brownian motion can emerge as solution of the Generalized Langevin Equation (GLE) 
\cite{GLE,GLE2,LEBook} for a Brownian
particle with mass $m$, 
\begin{equation}\label{GLE}
m\ddot x+\int_{0}^{t}\eta(t-t')\dot x(t')dt'=f(x,t)+\xi(t),
\end{equation}
where $f(x,t)=-\partial V(x,t)/\partial x$ is a deterministic force, 
$\xi(t)$ is zero-mean and Gaussian-distributed thermal noise and $\eta(t)$ is the frictional memory kernel related 
to noise  by the fluctuation-dissipation relation (FDR)
\begin{equation}\label{FDR}
\langle \xi(t')\xi(t)\rangle =k_B T\eta(|t-t'|)\;.
\end{equation}
The fBm emerges in the overdamped limit, $m\to 0$, of a force-free motion, $f\to 0$, for a power-law frictional kernel
$\eta(t) = \eta_{\alpha}t^{-\alpha}/\Gamma(1-\alpha)$,
with $0<\alpha<1$ [$\Gamma(x)$ is standard gamma-function], 
and the FDR-related noise $\xi(t)$, which is termed  the fractional Gaussian noise (fGN) \cite{Mandelbrot}. 
The corresponding GLE is also termed the fractional Langevin equation (FLE) 
\cite{FLE,FLE2,LEBook}. The GLE can
be derived from a standard Hamiltonian model of the Brownian motion based on the coupling of Brownian
particle to a thermal bath of harmonic oscillators modeling the environment \cite{GLE2}. The FLE model corresponds to 
sub-Ohmic thermal bath characterized by the spectral bath density 
$J(\omega)\propto \eta_{\alpha} \omega^{\alpha}$\cite{WeissBook}. 
Such a modeling requires a dense spectrum of the thermal bath, or \textit{quasi-infinite} number of oscillators.
Alternatively, one can model the environment by a \textit{finite} set of co-moving auxiliary Brownian particles,
cf. Fig. \ref{Fig1},
with masses $m_i$ coupled elastically with constants $k_i$ to the central Brownian particle and experiencing the
viscous Stokes friction (with frictional constants $\eta_i$) and uncorrelated, 
$\langle \xi_i(t)\xi_j(t')\rangle=\delta_{ij}\delta(t-t')$, white-noise thermal Gaussian forces 
$\sqrt{2\eta_ik_BT}\xi_i(t)$ \cite{Add}. Considering the overdamped limit for auxiliary particles, $m_i\to 0$, 
this yields upon introduction of fluctuating viscoelastic
forces, $u_i=-k_i(x-x_i)$ \cite{Add}:
\begin{eqnarray}\label{embedding1}
m\ddot x & =& f(x,t)+
\sum_{i=1}^{N}u_i(t) \;,\nonumber \\
\dot u_i& = &-k_i v-\nu_iu_i+\sqrt{2\nu_i k_i k_BT}\xi_i(t) \;,
 \end{eqnarray} 
where $\nu_i=k_i/\eta_i$ are the relaxation rates of viscoelastic forces. 
The last equation for $u_i$ has the form of relaxation equation for elastic force introduced by Maxwell in his
macroscopic theory of viscoelasticity \cite{Maxwell} which is augmented by the corresponding Langevin force 
in accordance with FDR. 
Such a description
was introduced in Refs. \cite{Goychuk09,Goychuk10} to model anomalous Brownian motion in complex viscoelastic media within
a generalized Maxwell model.  Indeed, by choosing initial $u_i(0)$ as independent  zero-mean Gaussian 
variables with variance $\langle u_i^2(0)\rangle=k_ik_B T$ and excluding the dynamics of auxiliary variables
$u_i$, the GLE (\ref{GLE}), (\ref{FDR}) follows immediately with the memory kernel  
$\eta(t)=\sum_{i=1}^{N}k_i e^{-\nu_i t}$.
For $N=1$, an earlier Markovian
embedding of GLE with exponentially decaying memory kernel \cite{Marchesoni} is readily reproduced. Furthermore, by
choosing $\nu_i=\nu_0/b^{i-1}, \;\; k_i=C_{\alpha}(b)\eta_{\alpha}\nu_i^{\alpha}/\Gamma(1-\alpha)$,
where $b$ is a scaling parameter and $C_{\alpha}(b)$ is  a fitting dimensionless 
constant, the  above power-law kernel $\eta(t)$ 
can be well approximated over about $r=N\log_{10}b-2$ temporal decades between two
time cutoffs, $\tau_{l}=\nu_0^{-1}<t<\tau_h=\tau_l b^{N-1}$. 
Similar scaling and approximation are well known in the anomalous relaxation theory \cite{Hughes,Palmer}.
Physically, $\nu_0$ corresponds to a high-frequency cutoff in $J(\omega)$, or the largest medium's frequency,
and $\nu_0/b^{N-1}$ to the slowest medium's mode, which are always present in reality.
For the case $\alpha=0.5$
studied numerically in this work, the choice $b=10$, $C_{0.5}=1.3$, $N=12$, and $\nu_0=100$ 
provides an excellent approximation  to the FLE dynamics over at least ten time decades, until $t_{\rm max}=10^8$.
This is checked \cite{Goychuk09,Goychuk10,SiegleEPL} by comparison with the exact solution 
for the position variance
obtained within and GLE and FLE \cite{GLE,FLE} in the force-free case. 
We scale time in units of the (anomalous) velocity relaxation constant
$\tau_v=(m/\eta_{\alpha})^{1/(2-\alpha)}$. It is assumed to
be temperature-independent in accordance with the Hamiltonian model \cite{GLE2,WeissBook}.
This is a standard assumption done also in other toy ratchet models \cite{ReimannRev,MarchesoniRev}. 

Stochastic dynamics is studied in a driven ratchet potential, $V(x,t)=U(x)-Ax\cos(\Omega t)+f_0 x$, where \cite{Bartussek}
\begin{eqnarray}
U(x)=-U_0[\sin(2\pi x/L)+0.25 \sin(4\pi x/L) ] 
\end{eqnarray}
is a spatially asymmetric periodic potential with amplitude $U_0$ and period $L$, $A$ and $\Omega$ are
the amplitude and frequency of the periodic forcing, and $f_0$ is a load.
The distance is scaled in the units of
$L$, the energy in units of $\tilde E=mL^2/\tau_v^2$, force in $\tilde F=\tilde E/L$ and 
temperature in $\tilde T=\tilde E/k_B$. The role of the inertial effects can be characterized by
the dimensionless parameter $r_v=1/(\omega_b\tau_v)$, where $\omega_b=(2\pi/L)(3^{3/8}/2^{1/4}) \sqrt{U_0/m} $ is the bottom 
and (imaginary) top frequency of the considered potential for $A=0$, $f_0=0$. The inertial effects can only be negligible 
for $r_v\ll 1$ and not too small $\alpha$ \cite{FLE2}. 
The corresponding borderline value of $U_0$ corresponding to $r_v = 1$ is
in the dimensionless units  $U_0^*\approx 0.0157$. For the simulations done in this work the inertial effects are very
essential. This might seem paradoxical since in the Markovian approximation $\dot x(t')\to\dot x(t)$ to Eq. (\ref{GLE})
the effective Markovian friction $\eta_{\rm eff}(t)=\int_0^{t}\eta(t')dt'\propto \eta_{\alpha}t^{1-\alpha}$ 
increases to infinity 
in the course of time. Such a Markovian 
approximation is, however, not affordable for the considered viscoelastic
dynamics. Fig. \ref{Fig1} illustrates this point: A part of auxiliary particles is strongly 
coupled to the Brownian particle.  They are co-moving being strongly correlated, reminding a polaron-like picture for quantum
particles in polar media. However, weaker
coupled, stronger damped, and much slower particles cannot follow immediately. They create an elastic force pulling the
central particle back and retarding its overall motion. This prohibits any Markovian approximation on the 
level of the reduced $(x,v)$ dynamics as the corresponding slow hidden dynamics cannot be adiabatically eliminated. 
And nevertheless, a highly dimensional Markovian approximation with $N$ 
extra dimensions for overdamped auxiliary particles works remarkably well. Of course, given a finite $N$ the trully
asymptotical dynamics becomes normal for $t\gg t_{\rm max}=b^{N-1}/\nu_{0}$. However, $t_{\rm max}$ grows exponentially
fast with $N$ and therefore it can be totally irrelevant, as in our simulations, to figure out the correct 
asymptotic FLE dynamics.  
We take a physical limit of very large $t$ to study subdiffusive dynamics 
with $t_{\rm max}$ regarded as infinite on this time scale and define the subvelocity as 
 $v_\alpha=\Gamma(1+\alpha)\lim\limits_{t\to\infty}\frac{\langle x(t)\rangle}{t^\alpha}$.
Practically, the corresponding values of $v_\alpha$ were calculated by fitting the dependence 
$\langle x(t)\rangle$ with the $at^\alpha$
dependence, extracting the corresponding $a$ within a last time window of simulations done until
$t_{\rm sim}=10^6\ll t_{\rm max}$. 
In all simulations we have used $n=10^4$ trajectories for ensemble averaging. The stochastic Heun algorithm has been implemented 
with double precision on a graphical processor unit (GPU) \cite{CUDA} which allowed to parallelize and accelerate 
stochastic simulations by a factor of about 50 as compare with standard CPU computing.  The amplitude of the 
external driving was fixed at 
$A=0.8$, which is beyond the lowest second order of nonlinear response in $A/T$ that yields a rectification subtransport 
response,
while the temperature, potential amplitude, and the driving frequency were varying.

\begin{figure}
 \includegraphics[width=\columnwidth]{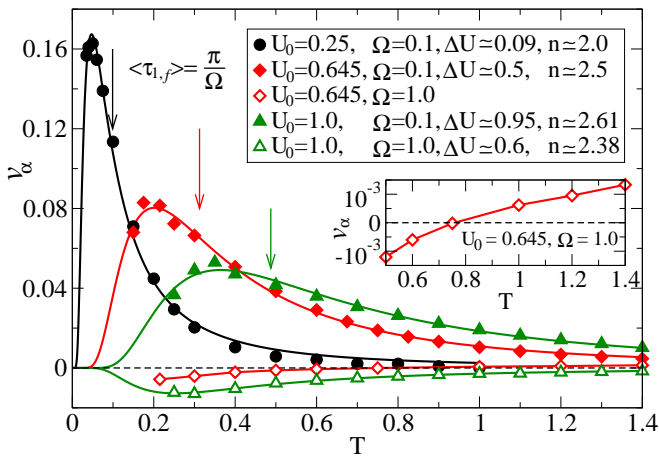}
 \caption{(Color online) Anomalous current (subvelocity $v_\alpha$) as a function of temperature for different values $U_0$ and $\Omega$.
          \label{Fig2}}
\end{figure}

\textit{Results.} First, we fixed the temperature at $T=0.25$ and varied the driving frequency and the potential amplitude, 
cf. \cite{Add}, Fig. S1.
Remarkably, the rectification response is absent in the adiabatic limit $\Omega\to 0$ as in \cite{Goychuk10}. 
This is in a sharp contrast with
the case of normal diffusion where the adiabatic current is maximal \cite{ReimannRev,MarchesoniRev}. 
Indeed, the subdiffusion and subtransport in 
periodic potentials turn out to be asymptotically insensitive  to the potential amplitude $U_0$ 
\cite{Goychuk09}. 
This surprising feature is due to the influence of slugish dynamics of the medium's degrees of freedom which 
cannot immediately
follow to a faster moving Brownian particle. They cause ultraslow dynamics on the time scale which largely exceeds
the mean time spent in a potential well. In this respect,  the medium's dynamics is not influenced directly by an external force. 
Therefore, for an adiabatically slow driving a periodic potential does not
play any role in the long time limit and the ratchet transport is absent. For a very fast driving,
the transport is also obviously increasingly suppressed
upon increasing the driving frequency.
Therefore, there should exist an optimal value of driving frequency when the corresponsing subvelocity  attains a maximum,
$v_{\alpha, max}=\max_{\Omega}v_{\alpha}(\Omega)$, which is a Stochastic Resonance related effect \cite{Add}. This maximal value is optimized also with the potential amplitude.   
For example, $v_{\alpha, max}$ is larger for $U_0=0.645$, than for $U_0=0.5$ and $U_0=1.0$, see 
in \cite{Add}, Fig. S1 and the corresponding
discussion therein.  Noteworthy is also the inversion of the current direction for sufficiently large $U_0$ and $\Omega$.
The subtransport then occurs in the conterintuitive direction, contrary to the direction predicted by the slow adiabatic tilt 
argumentation. A similar inversion for high-frequency driving occurs also in the normal diffusion case 
\cite{Bartussek,ReimannRev}.

SR-related phenomenon occurs also in the dependence of subcurrent on temperature, cf. Fig. \ref{Fig2}. 
An adiabatic driving argumentation predicts  
$v_\alpha(T)\propto T^{-n}\exp(-\Delta U/T)$ with $n=2$ and $\Delta U=2U_0$ in the lowest order of perturbation theory over
$A/T$ (quadratic response), for a weak driving, $AL\ll 2\pi U_0$. 
This is a typical SR dependence yielding the maximum versus temperature
at $T_{\rm max}=\Delta U/n$. We are dealing but with a strong nonadiabatic driving and hence use  
$\Delta U$ and $n$ in Fig. \ref{Fig2} as some fitting parameters.  
The SR origin of the maxima for sufficiently small $\Omega$ is substantiated by
matching the forward tilt half-periods with the numerical mean times $\langle \tau_{1,f}\rangle$ of the jumps 
into the forward directions, i.e. $\langle \tau_{1,f}\rangle=\pi/\Omega$ \cite{SR}, with temperature
near to the transport optimization
as indicated by arrows in Fig. \ref{Fig2}.   
 The negative current for a larger $\Omega$ is
also optimized with temperature, which does not have but any relation to SR or synchronization. 
Morever, an inversion of the subcurrent direction with temperature is detected
in Fig. \ref{Fig2} (see also the insert therein) for $U_0=0.645$ and a high-frequency driving $\Omega=1.0$. Similar temperature inversions occur also in the 
normal diffusion case \cite{Bartussek}.

Furthermore, the viscoelastic subdiffusion in periodic potentials is asymptotically also not affected by the potential height \cite{Goychuk09} 
even though a periodic driving can slightly affect it. Asymptotically, in the considered units 
$\langle \delta x^2(t)\rangle\sim 2 D_{\alpha} t^{\alpha}/\Gamma(1+\alpha)$ with subdiffusion coefficient 
$D_{\alpha}=T$. A periodic driving does affect 
$D_{\alpha}$. However, it is not changed strongly \cite{Goychuk10}.
Therefore, the generalized Peclet number\cite{Goychuk10}
${\rm Pe}_{\alpha}:=v_{\alpha}L/D_{\alpha}$, which measures the coherence and quality of transport \cite{Lindner}, can be
appreciable. 
With lowering the barrier height in Fig. \ref{Fig2} the maximum of subvelocity is shifted towards smaller temperatures and the
corresponding  ${\rm Pe}_{\alpha}$ can exceed the value of one signifying thereby that a high-quality subdiffusive ratchet 
transport is possible.

\begin{figure}
 \includegraphics[width=\columnwidth]{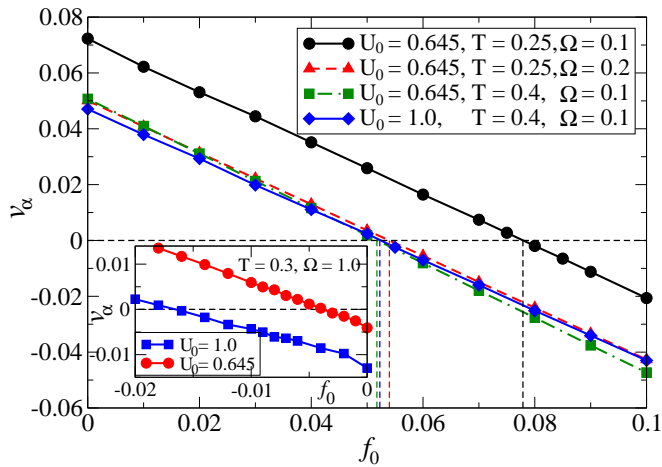}
 \caption{(Color online) Anomalous current (subvelocity $v_\alpha$) as a function of positive load $f_0$ 
          at different $U_0$, $T$ and a small driving frequency $\Omega=0.1$. Insertion shows dependencies $v_\alpha$
          versus negative load $f_0$ at $T=0.3$ and some high $\Omega=1.0$ at different $U_0$.
          \label{Fig3}}
\end{figure}

Finally, the dependence of the directed subtransport on the load $f_0$ in the opposite direction is shown in Fig. \ref{Fig3}. The existence
of the stopping force shows clearly that we are dealing with a genuine ratchet effect. Given the asymptotic 
independence of the viscoelastic subtransport on the potential amplitude $U_0$ in washboard potentials \cite{Goychuk09} one
expects a very simple dependence $v_{\alpha}(f_0)=v_{\alpha}(0)-f_0$ (in dimesionless units) to hold for small driving 
frequencies $\Omega$. Indeed,
the numerical results are consistent (taking numerical errors into account) with this prediction. However, for larger $\Omega$ 
and the inverted transport the deviations become more pronounced.

In conclusion, with work we put forward a toy model for viscoelastic subdiffusive ratchet transport and 
showed that such a transport 
can be optimized by ambient thermal noise and frequency of the external driving due to Stochastic Resonance effects and 
can exhibit a surprisingly good quality.  

\textit{Acknowledgment}. Support of this research by the Deutsche Forschungsgemeinschaft, Grant
GO 2052/1-1 is gratefully acknowledged.

\begin{eqnarray}\label{embedding2}
m\ddot x & =& f(x,t)-
\sum_{i=1}^{N}k_i(x-x_i) \;,\nonumber \\
m_i\ddot x_i& = &k_i (x-x_i)-\eta_i\dot x_i+\sqrt{2\eta_ik_BT}\xi_i(t) \;.
 \end{eqnarray}
This equations are similar to the starting equations of motion in the
purely dynamical model leading to the GLE description. 
The last one can be obtained by setting $\eta_i\to 0$, i.e.
considering a purely dynamical evolution with a tremendously large
number of oscillators $N\to \infty$ with a dense spectrum $\omega_i=\sqrt{k_i/m_i}$. 
The only 
nondynamical assymption in the dynamical model is that \textit{initially}
the thermal bath oscillators are canonically distributed at the 
temperature $T$ -- like in a standard molecular dynamics setup. We replace them 
by a \textit{finite} set of auxiliary 
Brownian particles.

\begin{figure}
\includegraphics[width=\columnwidth]{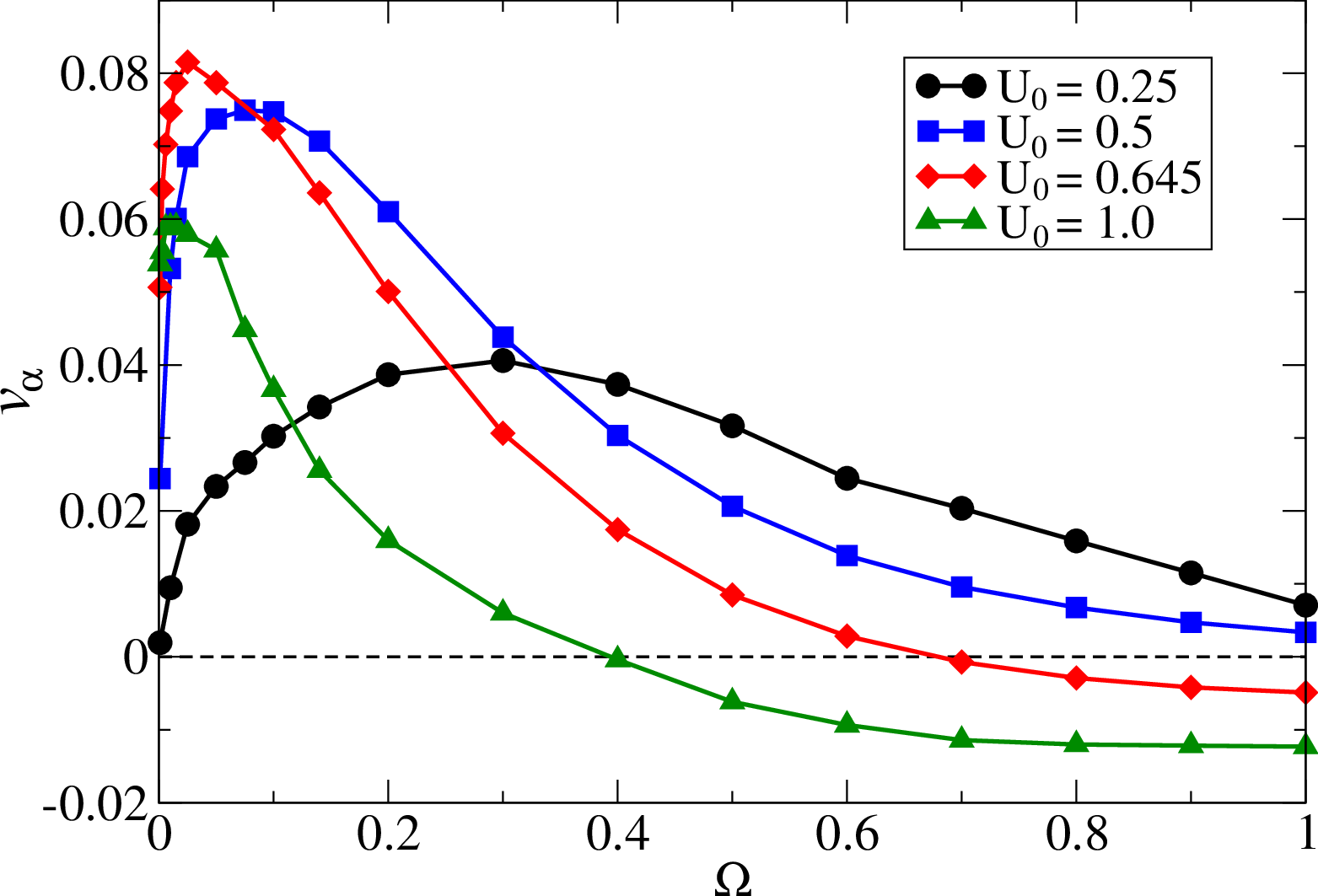}
Figure S1: Anomalous current (subvelocity $v_\alpha$) as a function of the driving frequency
for different $U_0$ at $T=0.25$ for zero load $f_0=0$. 
\label{FigS1}
\end{figure}

\begin{figure}
\includegraphics[width=\columnwidth]{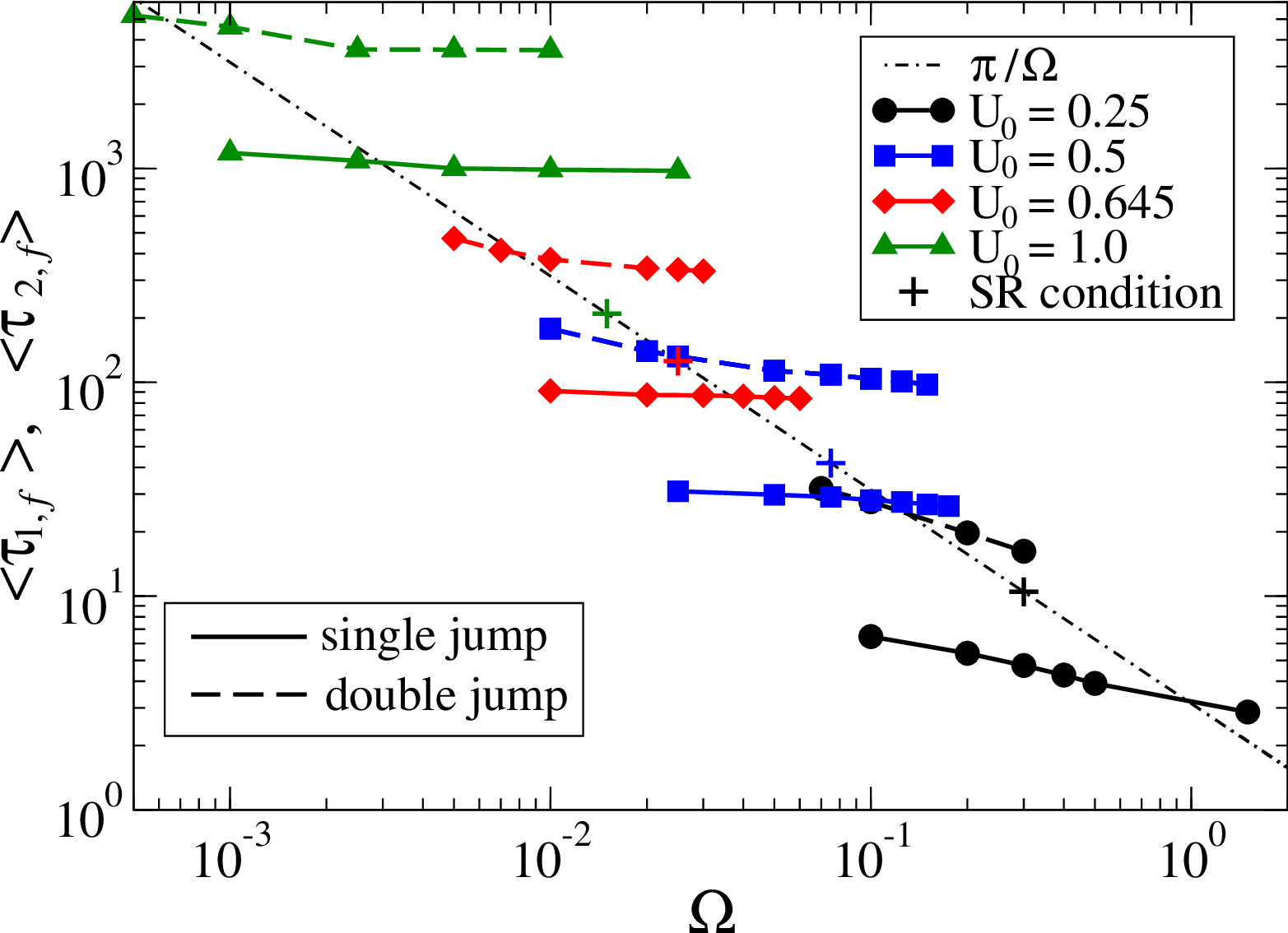}
Figure S2: Mean times of transitions, $\langle \tau_{1,f}\rangle$ and $\langle \tau_{2,f}\rangle$, 
as functions of the driving frequency $\Omega$. The values $\Omega_{\rm max}$ and  
$v_{\alpha}(\Omega_{\rm max})$ are indicated by crosses.
\label{FigS2}
\end{figure}

\textbf{2.} Fig. S1 displays the dependence of subvelocity on external driving frequency $\Omega$ for different potential
amplitudes $U_0$ and the fixed driving amplitude and temperature. In order to clarify
the physical origin of maximal $v_{\alpha, max}$  we have calculated the frequency-dependent mean times of the 
transitions to the neighboring potential well, $\langle \tau_{1,f}\rangle$, and to the second next potential well,
$\langle \tau_{2,f}\rangle$. If the maximum of $v_{\alpha}$ is due to synchronization, then the transport should become
optimal when the particles advance in the trasport direction over one or two potential periods during the 
driving half-period ${\cal T}_{1/2}=\pi/\Omega$. This means that the condition 
$\langle \tau_{1,f}\rangle<{\cal T}_{1/2}<\langle \tau_{2,f}\rangle$ should be obeyed. Indeed, the numerical results
displayed
in Fig. S2 show
that this is the case for not too high $U_0$. However, since $\langle \tau_{1,f}\rangle$ and $\langle \tau_{2,f}\rangle$
are only weakly frequency-dependent and thus no frequency entraintment occurs, we are dealing with the
phenomenon of Stochastic Resonance rather than stochastic synchronization. 
Here, the response maximizes when an external driving frequency fits into an intrinsic
frequency of stochastic dynamics.  

\end{document}